\documentclass[aps,prd,twocolumn,showpacs,superscriptaddress,footinbib]{revtex4-1}  
\usepackage{graphicx}  
\usepackage{dcolumn}   
\usepackage{bm}        
\usepackage{comment}
\usepackage{amssymb}   
\usepackage{amsmath}
\usepackage[usenames,dvipsnames]{color}
\usepackage[normalem]{ulem}
\usepackage{soul}

\definecolor{orange}{RGB}{255,127,0}

\setstcolor{orange}

\definecolor{blue}{RGB}{255,127,0}

\setstcolor{blue}

\usepackage{graphicx}
\usepackage{dcolumn}
\usepackage{bm}
\usepackage{comment}
\usepackage{dcolumn}   
\usepackage{amssymb}   
\usepackage{amsmath}
\usepackage{tabularx}
\usepackage{hyperref}
\usepackage{xcolor}
\usepackage{tabulary}

\makeatletter
\def\mathcolor#1#{\@mathcolor{#1}}
\def\@mathcolor#1#2#3{%
  \protect\leavevmode
  \begingroup
    \color#1{#2}#3%
  \endgroupt
}
\makeatother

\begin{document}

\preprint{APS/123-QED}

  \title{Black Holes: The Next Generation -- Repeated Mergers in Dense Star Clusters and their Gravitational-Wave Properties}

  \author{Carl L.\ Rodriguez}
  \affiliation{MIT-Kavli Institute for Astrophysics and Space Research, 77 Massachusetts Avenue, 37-664H, Cambridge, MA 02139, USA}
  
              \author{Michael Zevin}
\affiliation{Center for Interdisciplinary Exploration and Research in
    Astrophysics (CIERA) and Dept.~of Physics and Astronomy, Northwestern
      University, 2145 Sheridan Rd, Evanston, IL 60208, USA}
      
       \author{Pau Amaro-Seoane}
          \affiliation{Institute of Space Sciences (ICE, CSIC) \& Institut d'Estudis Espacials de Catalunya (IEEC)\\
at Campus UAB, Carrer de Can Magrans s/n 08193 Barcelona, Spain\\
Kavli Institute for Astronomy and Astrophysics, Beijing 100871, China\\
Institute of Applied Mathematics, Academy of Mathematics and Systems Science, CAS, Beijing 100190, China\\
Zentrum f{\"u}r Astronomie und Astrophysik, TU Berlin, Hardenbergstra{\ss}e 36, 10623 Berlin, Germany}

  \author{Sourav Chatterjee}
         \affiliation{Tata Institute of Fundamental Research, Department of Astronomy and Astrophysics, Homi Bhabha Road, Navy Nagar, Colaba, Mumbai, 400005, India}

  \author{Kyle Kremer}
\affiliation{Center for Interdisciplinary Exploration and Research in
    Astrophysics (CIERA) and Dept.~of Physics and Astronomy, Northwestern
      University, 2145 Sheridan Rd, Evanston, IL 60208, USA}

        \author{Frederic A.\ Rasio}
\affiliation{Center for Interdisciplinary Exploration and Research in
    Astrophysics (CIERA) and Dept.~of Physics and Astronomy, Northwestern
      University, 2145 Sheridan Rd, Evanston, IL 60208, USA}

        \author{Claire S.\ Ye}
\affiliation{Center for Interdisciplinary Exploration and Research in
    Astrophysics (CIERA) and Dept.~of Physics and Astronomy, Northwestern
      University, 2145 Sheridan Rd, Evanston, IL 60208, USA}

\date{\today}


\begin{abstract}
When two black holes merge in a dense star cluster, they form a new black hole with a well-defined mass and spin.  If that ``second-generation'' black hole remains in the cluster, it will continue to participate in dynamical encounters, form binaries, and potentially merge again.
Using a grid of 96 dynamical models of dense star clusters and a cosmological 
model of cluster formation, we explore the production of binary black hole 
mergers where at least one component of the binary was forged in a previous 
merger.   We create four hypothetical universes where every black hole born in 
the collapse of a massive star has a dimensionless Kerr spin parameter, 
$\chi_{\rm birth}$, of 0.0, 0.1, 0.2, or 0.5.  We show that if all stellar-born 
black holes are non-spinning ($\chi_{\rm birth}=0.0$), then more than 10\% of 
merging binary black holes from clusters have components formed from previous 
mergers, accounting for more than 20\% of the mergers from globular clusters 
detectable by LIGO/Virgo.  Furthermore, nearly 7\% of detectable mergers would 
have a component with a mass $\gtrsim 55M_{\odot}$, placing it clearly in the mass ``gap'' region where black holes cannot form from isolated collapsing stars due to the pulsational-pair instability mechanism.   On the other hand, if black holes are born spinning, then the contribution from these second-generation mergers decreases, making up as little as 1\% of all detections from globular clusters when $\chi_{\rm birth} = 0.5$.  We make quantitative predictions for the detected masses, mass ratios, and spin properties of first- and second-generation mergers from dense star clusters, and show how these distributions are highly sensitive to the birth spins of black holes.

\end{abstract}

\maketitle

\section{Introduction}
\label{sec:intro}

As of 2019, the majority of detected stellar-mass black 
holes (BHs) have been detected through gravitational waves (GWs).
The first two observing runs of LIGO and Virgo (O1 and O2) yielded 10 binary
black hole (BBH) mergers \cite{Abbott2016,Abbott2017,Abbott2017e,Abbott2017c,Abbott2016a,TheLIGOScientificCollaboration2018a}, while the 
ongoing O3 run has already reported several significant BBH candidates.  
Before the decade is complete, we will likely have information about the masses, 
spins, and cosmological redshifts of more than 100 BHs. While there exist many proposed mechanisms for forming double compact object mergers, such as the evolution of massive binary stars \cite{Podsiadlowski2003,Dominik2012,Dominik2014,Dominik2013,Belczynski2016,Voss2003}, 
dynamical formation in dense star clusters \cite{Askar2016,Tanikawa2013,Hong2018,Downing2011,OLeary2006,Moody2009,Bae2014,PortegiesZwart2000,Choksi2018a,Chatterjee2017,Downing2010,Rodriguez2016a,Rodriguez2016b,Rodriguez2015a}, long-term 
secular interactions with one (or more) bound companions 
\cite{Hoang2018,Prodan2014,Antognini2014,Antonini2012,Antonini2015,Antonini2017,Silsbee2017,2018MNRAS.480L..58A,2017ApJ...846L..11L,2018ApJ...863....7R,2019ApJ...878...58S,2019arXiv190500427L,2019MNRAS.486.4781F,2019MNRAS.488...47F}, migration and capture in 
AGN disks \cite{Stone2016,Bartos2016,McKernan2018,Secunda2018}, and even formation from primordial BHs 
\cite{Bird2016}, the vast majority of these 
formation channels source their component BHs from the collapse of massive 
stars.  The outcome of stellar collapse should obey similar physics regardless 
of the formation channel or merger environment.  

When a BBH merges in isolation, the resultant BH is unlikely to interact 
again with other stars or BHs.  But when a merger occurs in a dense stellar 
environment, such as a globular cluster (GC) or nuclear star cluster (NSC), the 
fate of the remnant can be far more interesting.  For many years, it was 
assumed that most BHs produced from the mergers of other BHs would be ejected 
from their host clusters 
\cite[][]{Merritt2004,2007ApJ...667L.133S,Gualandris2008,HolleyBockelmann2008}, because when the 
spins of the BBH components are large, the merger products receive large 
kicks $(\sim 10^3 \rm{km}/\rm{s})$ due to the asymmetric emission of GWs \cite{2007PhRvL..98w1101G,2007PhRvL..98w1102C}.  However, GW observations have 
suggested that many of the BBH mergers observed by LIGO/Virgo may have involved BHs with low intrinsic spin, significantly reducing the recoil kicks 
experienced by the merger products \cite{TheLIGOScientificCollaboration2018a,TheLIGOScientificCollaboration2018}.

If the recoil velocity of the merging binary is less than the local escape 
speed, the newly-formed BH will be retained by the cluster, creating a new 
generation of BHs.  These second-generation (2G) BHs will continue to 
participate in three- and four-body dynamical encounters, eventually forming new BBHs and potentially merging a second time \cite[][]{Antonini2016,Rodriguez2018,Antonini2019}.  These  
mergers have unique masses, mass ratios,
and spins which may be difficult or impossible to produce from first-generation
(1G) BBHs produced from collapsing stars \cite{Gerosa2017,Fishbach2017}.  In particular, both theoretical modeling of massive 
stars \cite{Woosley2015,Woosley2017,Woosley2019} and statistical modeling of the LIGO/Virgo BBH catalog 
\cite{Fishbach2017a,TheLIGOScientificCollaboration2018,2018ApJ...856..173T} have suggested the existence of a gap in the BH 
mass function above $\sim 40M_{\odot}$, arising from pulsational pair 
instabilities (PPIs) and pair-instability supernovae (PISN).  
The detection of BHs in this upper-mass gap would be strong evidence 
for the dynamical processing of BHs prior to their eventual merger.

In this paper, we explore the properties of 1G and 2G BBH mergers created from a 
realistic collection of GC models.  Using a cosmological model for star-cluster 
formation, we create four hypothetical universes where the birth spins of 1G 
BHs, $\chi_{\rm birth}$, are uniformly 0.0, 0.1, 0.2, or 0.5.
As the birth 
spin of the BHs is increased, the retention of the BBHs that merge in the cluster decreases, changing the mass and spin distributions of the BBHs detectable by LIGO/Virgo.  In Section \ref{sec:method}, we describe the physics of our GC models, and the weighting scheme we use to reproduce the cosmological formation/evolution of GCs and the detectable population of BBHs.  In Section \ref{sec:2gbbh}, we show how the retention of 2G BHs depends on the birth spins.  We also describe what fraction of BBH mergers may contain a 2G BH, and what fraction of those sources would lie in the PPI/PISN mass gap.   In Section \ref{sec:massspin}, we show the mass, mass ratio, and spin distributions of all 1G and 2G BBHs, and compare them to the current catalog of GW observations.  Throughout this paper, we assume a flat $\Lambda \rm{CDM}$ cosmology with $h= 0.679$ and $\Omega_M = 0.3065$
\cite{PlanckCollaboration2015}. We describe the composition of BBHs by the 
generation of their components (e.g.~a 2G+1G BBH has one 2G component and one 1G 
component), where 1G BHs are created from collapsing stars, and 2G components 
are created in a previous BBH mergers.

\section{Methods}
\label{sec:method}

We generate 96 models of dense star clusters using the \texttt{Cluster Monte 
Carlo} (CMC) Code, a H\'enon-style $N$-body code for stellar dynamics \citep{Joshi1999,Pattabiraman2013}.  Because the H\'enon Monte Carlo approach can model the dynamics of individual stars in a cluster, CMC can explicitly follow the formation and evolution of potential GW sources over many Gyr.  The stars and binaries in our models are evolved self-consistently from their zero-age main-sequence births using the binary stellar evolution (BSE) package \cite{Hurley2000,Hurley2002,Chatterjee2010}, with updated prescriptions for the formation of BHs and NSs from massive stars \citep[][and references therein]{Rodriguez2016a} and the PPI/PISN physics \cite{Rodriguez2018}. These stars and binaries move dynamically through the cluster, where they participate in all the gravitational dynamics --- collisional diffusion following the Fokker-Planck approximation \citep{Henon1975,Joshi1999}, binary formation in three-body encounters \cite{Morscher2012}, strong gravitational encounters between stars and binaries \cite{Fregeau2007}, tidal stripping by the galactic potential --- that can form merging BBHs.  

\subsection{Spins, Kicks, and post-Newtonian Dynamics}
\label{subsec:model}

In \cite{Rodriguez2018}, we added post-Newtonian (pN) corrections to the orbital dynamics of isolated 
BBHs and strong gravitational encounters involving BHs using the code developed and tested in \cite{Antognini2014,Amaro-Seoane2016}.
For BBHs 
that merge inside the cluster, we self-consistently calculate the final mass, 
spin, and recoil velocity of the BH merger product using detailed fitting 
formula from numerical relativity simulations \cite{Merritt2004,Berti2007,Campanelli2007b,Gonzalez2007,Kesden2008,Tichy2008,Buonanno2008,Rezzolla2008,Lousto2008,Barausse2009,Barausse2012,Lousto2012,Lousto2013,Lousto2014}.  See \cite{Rodriguez2018}, Appendix A for a complete description of the equations used.  However, in 
that study we only considered 1G BHs with zero spin (although we extrapolated our results to higher birth 
spins).  Furthermore, we later showed in \cite{Rodriguez2018c} 
that naively including the first and second 
pN corrections to the equations-of-motion can introduce significant biases in 
the measured eccentricities and binary classifications during strong encounters.  

In this paper, we perform the first self-consistent estimates of 2G BBH 
formation with varying initial BH spins and a realistic model for GC formation.  Our GC initial conditions are 
identical to those presented in \cite{Rodriguez2018c}, and cover a 
range of initial particle numbers ($2\times10^5$, $5\times10^5$, $10^6$, and 
$2\times10^6$), initial virial radii (1pc and 2pc), and galactocentric 
radii/stellar metallicities (2 kpc/$0.25Z_{\odot}$, 8 kpc/$0.05Z_{\odot}$, and 
20 kpc/$0.01Z_{\odot}$).  In this study, we expand this grid in a fourth 
dimension, and consider initial 1G BH spins of 0.0, 0.1, 0.2, and 
0.5, for a total of 96 GC models; although the spins of BHs can go as high as 1, we find that a maximum $\chi_{\rm birth}$ of 0.5 is already sufficient to eject the vast majority of 1G+1G mergers from the cluster.  We do not consider more complicated prescriptions for BH spin based 
on the stellar mass and metallicity, such as those presented in 
\cite{Belczynski2017}.  However, those results suggest that heavier BHs ($\sim 
30M_{\odot}$) may be born with spins in the 0.0-0.2 range, allowing us to extrapolate 
from the models presented here.  The initial positions and velocities of individual particles are drawn from a
King profile \cite{King1966} with a concentration of $w_0=5$.  The initial stellar masses are chosen from a Kroupa initial mass function (IMF) \cite{Kroupa2003} in a range between $0.08 M_{\odot}$ to $150M_{\odot}$.  
We assume that 10\% of objects are initially in binaries, with semi-major axes distributed flatly in log from the point of stellar contact to the local hard/soft boundary.  Binary eccentricities are drawn from a thermal distribution, $p(e)de = 2e~de$.  The primary mass, $m_1$, of each binary is taken from the IMF, while the secondary mass is drawn from a flat distribution from 0 to $m_1$.

\subsection{GC Population and Detection Weights}
\label{sec:mccosmo}

In \cite{Rodriguez2018}, we presented results from a series of GC models with a subset of the initial conditions presented here, but with no differentiation between clusters of different masses, ages, and metallicities.  Here, we draw our BBH samples from each GC model according to a cosmologically-motivated model for GC formation 
\citep{El-Badry2018}.  This 
model was first used to predict the merger rate of BBHs from GCs in 
\cite{Rodriguez2018b}, and the weighting scheme we use is described in detail in 
\cite{Rodriguez2018c}.  Briefly, this procedure assigns to each GC model a weight based on 
how often clusters of that mass and metallicity are formed in the semi-analytic 
model of \citep{El-Badry2018}.  We divide the masses of GCs into 4
logarithmically-spaced bins, with one GC model in the center of each bin.  The 
cluster models are then assigned a weight according to the integral of the cluster initial 
mass function (CIMF) over the extent of that bin.  We assume a CIMF proportional to 
$1/M_{\rm GC}^2$; although there is evidence that the CIMF may contain an 
exponential-like truncation at higher masses \cite[][and references 
therein]{PortegiesZwart2010}, we find that our results are largely 
insensitive to such a choice \cite[though the same cannot be said for the 
overall BBH merger rate; see][]{Rodriguez2018b}.  We also divide the 
metallicity of GCs into three bins, and assign each cluster a weight based on 
the fraction of GCs formed at that metallicity at that redshift (using the median star-formation 
metallicity in a given halo mass at a given redshift from 
\cite{Behroozi2013} and the relation between stellar and gas metallicity 
from \cite{Ma2016}).  The weight assigned to each GC model is the product of the 
mass and metallicity weights.  For each BBH merger, we convolve 
the merger time of the BBH with the distribution of formation times for GCs of that metallicity 
\cite[See][Figure 1]{Rodriguez2018c} by drawing 100 random GC formation times 
for that BBH, and adding each merger to our sample.  In other words, the merger time of a BBH is the cosmic time when that GC formed plus the time taken for the BBH to form and merge in that cluster.  BBHs that merge later than the 
present day are discarded.  Each BBH is then assigned the weight associated with 
its parent cluster, and it is these weights we use to create the results
presented here.  We note that our cluster formation model encompasses dense stellar clusters beyond the classical GCs observed in the local universe, and also includes low-mass open clusters which have disrupted before the present day, as well as super-star cluster formation in the local universe.  However, we refer to this population as GCs for simplicity (and since the majority of BBH production occurs in these massive, old systems).

This weighting procedure provides us with the underlying physical distribution of sources at a given redshift interval per comoving volume, and throughout this paper we present results over all redshifts and in the local universe (defined as $z<1$).  But we are also interested in the distribution of sources that can be detected by LIGO/Virgo, for which we must consider both the increased sensitivity of the detectors to BBHs of higher masses and the larger amount of comoving volume surveyed at higher redshifts.  To that end, we also report a detectable distribution of BBH mergers, created by multiplying the astrophysical weights by an additional detectability weight.  That weight is calculated with:

\begin{equation}
    w_{\rm det} \equiv f_d (m_1,m_2,\vec{\chi}_1,\vec{\chi}_2,z) \frac{dVc}{dz} \frac{dt_s}{dt_o} ~,
\end{equation}

\noindent where $f_d(m_1,m_2,\vec{\chi}_1,\vec{\chi}_2,z)$ is the fraction of sources with masses $m_1$ and $m_2$ and spin vectors $\vec{\chi}_1$ and $\vec{\chi}_2$ merging at redshift $z$ that are detectable by LIGO/Virgo, $\frac{dVc}{dz}$ is the comoving volume at a given redshift, and $\frac{dt_s}{dt_o} = 1/(1+z)$ is the time dilation between clocks at the source and clocks on Earth. 

To calculate $f_d(m_1,m_2,\vec{\chi}_1,\vec{\chi}_2,z)$, we first determine 
$\vec{\chi}_1$ and $\vec{\chi}_2$ by randomly drawing the spin angles 
isotropically on the sphere. 
We calculate the optimal matched-filter signal-to-noise ratio ($\rm SNR$), $\rho_{\rm opt}$, for each sample using a 3-detector network configuration consisting of the Hanford, Livingston, and Virgo interferometers with projected design sensitivities \cite{Abbott2018}. Waveforms are generated using the IMRPhenomPv2 approximant \cite{Hannam2014}. 
Using an $\rm SNR$ detection threshold of $\rho_{\rm thresh} = 8.0$, if $\rho_{\rm opt} < \rho_{\rm thresh}$, the system is undetectable and $f_d(m_1,m_2,\vec{\chi}_1,\vec{\chi}_2,z) = 0$. 
Otherwise, we randomly sample the sky location, inclination angle, and polarization angle of each potentially detectable system $N = 10^4$ times and calculate the $\rm SNR$, $\rho_i$, for each of these realizations.
$f_d(m_1,m_2,\vec{\chi}_1,\vec{\chi}_2,z)$ is the fraction of these systems that exceed $\rho_{\rm thresh}$: 
\begin{equation}
    f_d(m_1,m_2,\vec{\chi}_1,\vec{\chi}_2,z) = \frac{1}{N}\sum_i^N \Theta(\rho_i - \rho_{\rm thresh})~, 
\end{equation}
where $\Theta$ is the Heaviside step function.

\section{Second Generation Black Holes}
\label{sec:2gbbh}

\subsection{Birth Spins and Black Hole Retention}

\begin{figure}[tb]
\centering
\includegraphics[scale=0.85, trim=0in 0.5in 0in 0.5in, clip=true ]{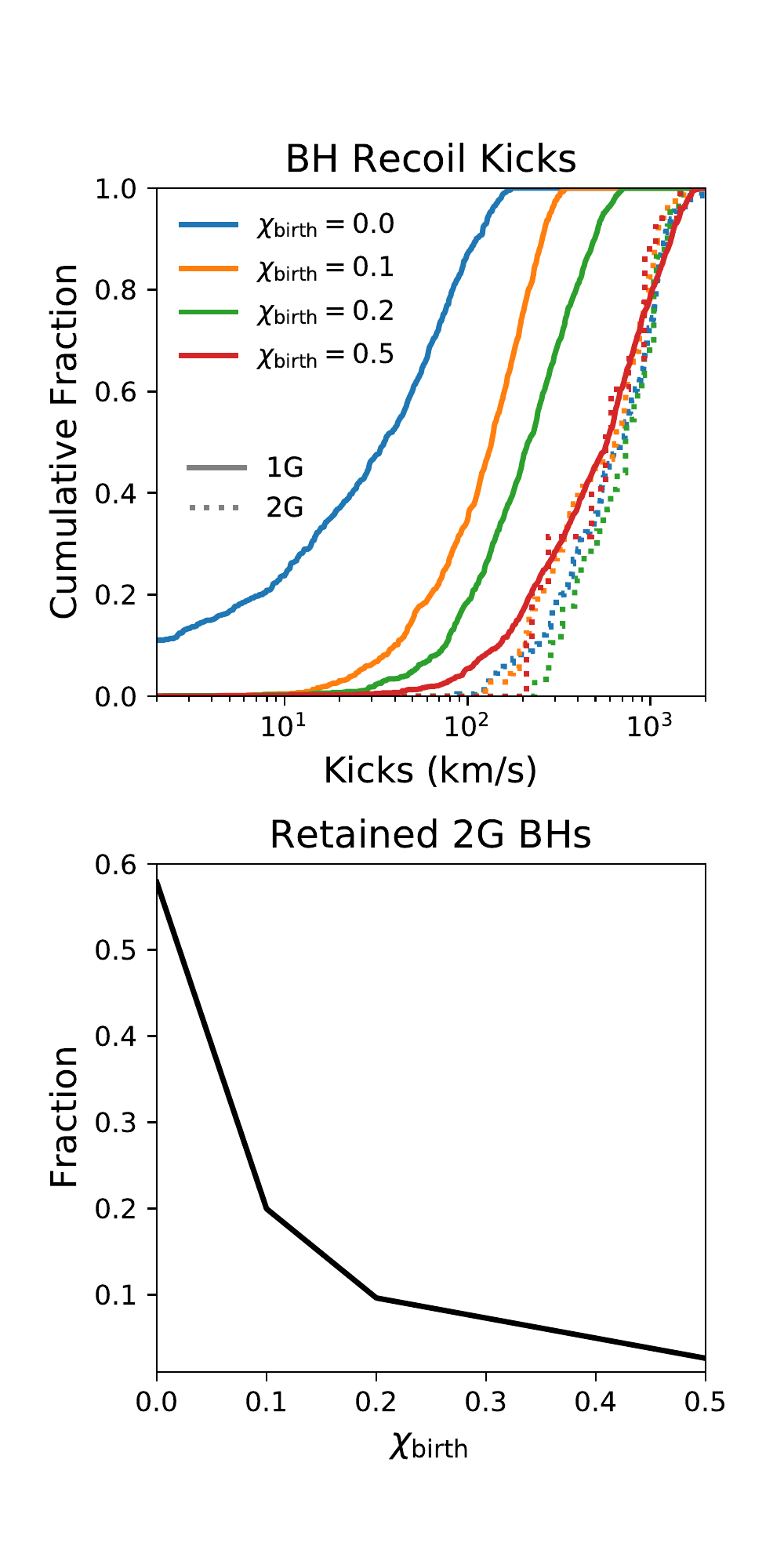}
\caption{The effect of initial BH spin on the recoil kicks and retention of BBH 
merger products.  On the top, we show the cumulative distribution of BH kicks for our $\chi_{\rm birth} = 0.0$, 0.1, 0.2, and 0.5 models in blue, orange, green, and red, respectively.  The solid lines show the distribution kicks for binaries comprised of both 1G components, while the dotted lines show the kicks for merging binaries that include at least one 2G BH.  On the 
bottom, we show the fraction of all BBH merger products that are retained in the 
cluster as a function of the birth spins of BHs.}
\label{fig:kicks}
\end{figure}

\begin{figure}[tb]
\centering
\includegraphics[scale=0.85, trim=0in 0.in 0in 0.in, clip=true ]{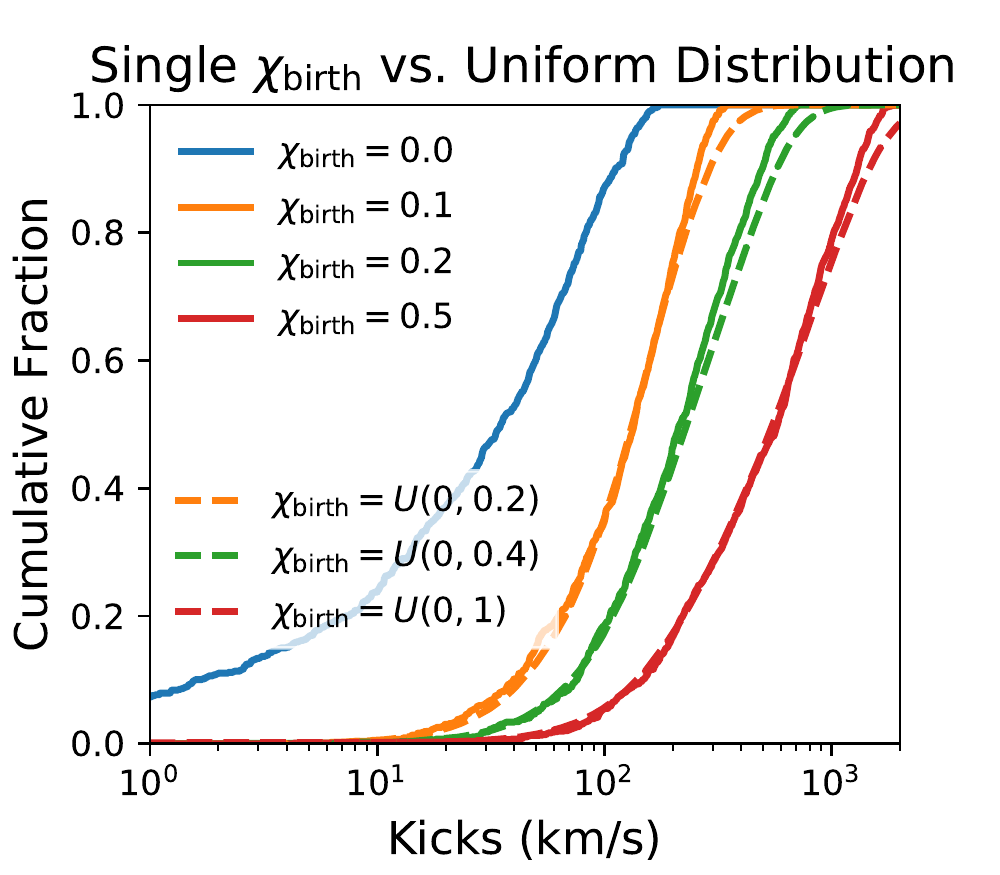}
\caption{Same as the top panel of Figure \ref{fig:kicks}, but showing the cumulative kick distribution from the four sets of spin models (solid lines), as well as the kick distributions the BBHs would have experienced if their spin magnitudes had been drawn from uniform distributions (dashed lines).  Our $\chi_{\rm birth}=0.1$, 0.2, and 0.5 models produce nearly identical kicks to models where the spins were drawn from uniform distributions between 0 and 0.2, 0.4 and 1, respectively.}
\label{fig:kick}
\end{figure}

The key question in the production of 2G BHs in GCs is whether the BBH mergers 
that occur in the cluster can be retained by the cluster.  In Figure 
\ref{fig:kicks}, we show the fraction of BBH merger products that are retained in 
their host clusters as a function of birth spin, $\chi_{\rm birth}$.  For the 
case where the birth spins of 1G BHs are zero, nearly 60\% of the merger 
products are retained in the cluster, since the typical GW kicks are typically 
limited to  $\lesssim$ 100 $\rm{km}/\rm{s}$, and depend entirely on the mass 
ratio of the system.  However, in the case where either component has 
significant spin, either from birth or a previous merger, then the kicks can 
exceed 1,000 $\rm{km}/\rm{s}$, significantly beyond the typical escape speeds of 
GCs.  In the $\chi_{\rm birth} = 0.5$ case, less than 3\% of merger products are 
retained.  Furthermore, because the merger of two BHs creates a new BH with a 
spin $\sim 0.7$, 1G+2G and 2G+2G BBH merger products are virtually never retained by the 
cluster, regardless of $\chi_{\rm birth}$.  Out of 96 GC models and nearly $10^4$ BBH mergers, we only identify one case where a 1G+2G merger is retained by the cluster, owing to the chance alignment of its spins.  That merger product forms another binary and is rapidly ejected from the cluster, creating a 3G+1G BBH merger.  We also identify one case where a merger takes place during a strong encounter between a single BH and a BBH.  The merger product is ejected from the cluster, but it remains bound to the third BH from the triple encounter, merging as a 3G+1G binary in the field \cite[the ``double mergers'' identified by][]{Samsing2019}.  For simplicity, we count these systems as 2G+1G BBHs in our results, as their component masses (both $\sim 80M_{\odot}+30M_{\odot}$) would not distinguish them as a 3G BH.  However, we do note that the spin magnitudes of these 3G BHs ($\chi \sim 0.39$ and $0.45$) are distinct from the 2G BHs, as is typical for mergers with small mass ratios \cite[e.g.,][]{Berti2007}.

\begin{figure}[tbh]
\centering
\includegraphics[scale=0.78, trim=0.1in 0.55in 0in 0.2in, clip=true ]{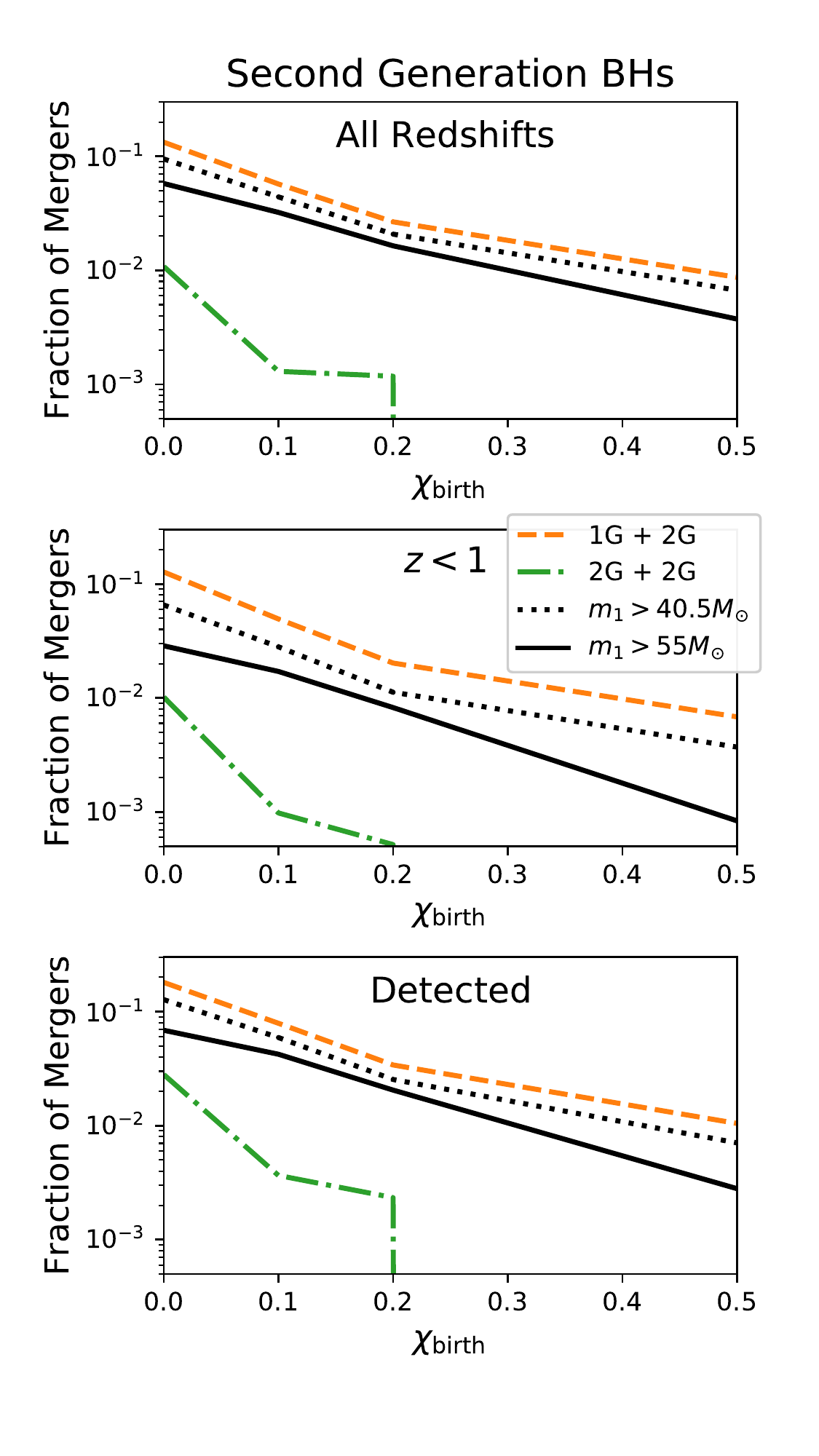}
\caption{The fraction of BBH mergers that are 
comprised of 1G and 2G BHs as a function of the birth spins of 1G BHs.  In 
dashed orange, we show the fraction of BBH mergers that contain one 2G BH, while 
dotted green shows the fraction of mergers with both 2G components.  We also 
show the fraction of mergers with at least one 2G component that is greater than 
$40.5M_{\odot}$ in dotted black (the beginning of the PPI/PISN mass gap in our 
stellar evolution prescriptions), and greater than $55M_{\odot}$ in solid black (our more conservative lower limit for the beginning of the mass gap).  The top panel shows the relative fraction of 2G mergers over all redshifts, while the middle panel shows the mergers in the local universe ($z<1$).  The bottom panel indicates the relative fraction of mergers of each type detectable by a three-detector LIGO/Virgo network operating at design sensitivity (see Section \ref{sec:mccosmo}).}
\label{fig:massgap}
\end{figure}


As noted above, our choice of birth spins for BHs -- setting $\chi_{\rm birth}$ to discrete values for all BHs regardless of the details of the pre-collapse star -- is highly simplistic.  In reality, it is entirely possible that the structure of the massive stars that form BHs can determine the final spin of the remnant, since the radial profile of the star will determine the efficiency of angular momentum transport out of the core once the star evolves onto the giant branch.  Recent detailed studies massive stars \cite{2019MNRAS.485.3661F,2019arXiv190703714F}, in particular, have suggested that the magnetic Taylor instability may be highly efficient at transporting angular momentum out of the cores of stars prior to collapse, birthing BHs with extremely low ($\chi_{\rm birth} \sim 0.01$) spins.  Unfortunately, to the authors knowledge there exist no published models relating stellar core masses, envelope masses, and metallicities to final BH spins for the range of initial conditions considered here.   Studies of the BBH population from LIGO/Virgo \cite[e.g.][]{TheLIGOScientificCollaboration2018,2018arXiv180506442W} have similarly been unable to constrain the BH spin magnitude distribution (although the possibility that all BHs are born with near-maximal spins is becoming disfavored).  

Our choice of discrete birth spins is different from previous dynamical studies in the literature \cite[e.g.,][]{Antonini2016}, which drew spin magnitudes from a distribution uniform between 0 and 1 (as well as a high-spin model, which is now disfavored).  While these choices are both arbitrary, they produce similar results.  The distribution of kicks for merging BBHs with spin magnitudes drawn from a $U(0,\chi_{\rm max})$ distribution is nearly identical to the distribution of kicks when $\chi_{\rm birth} = \chi_{\rm max} / 2$, when marginalizing over all BBH spin orientations and magnitudes.  In Figure \ref{fig:kick} we show the kick distributions from Figure \ref{fig:kicks}, as well as the distributions of kicks the merging BHs would have received if their spin magnitudes had been drawn from uniform distributions.  The distributions are virtually identical except at large kick velocities (where in both cases the binaries would be ejected from the cluster).  This makes our $\chi_{\rm birth} = 0.1$, 0.2, and 0.5 models equivalent to models where the spin magnitudes are drawn uniformly from 0 to 0.2, 0 to 0.4, and 0 to 1, respectively.

\subsection{Black Holes in the Upper Mass Gap}

The easiest identifying feature of 2G BHs is their characteristically large mass.  Both theoretical considerations and the first few observed LIGO BBH mergers have suggested the presence of an upper mass gap of stellar-born BHs, where compact objects cannot form due to PPIs/PISNs.  In a sufficiently massive, post-carbon burning star with a helium core mass $\gtrsim 30M_{\odot}$, the conversion of photons to electron-positron pairs removes pressure support from the core on a dynamical timescale.  In response, the stellar core contracts rapidly, increasing to temperatures sufficient for carbon, oxygen, and silicon burning \cite[e.g.][]{Woosley2015}.  If  this injection of energy is less than the binding energy of the star, as is the case for helium-core masses in the $30M_{\odot}-64M_{\odot}$ range, then these PPIs will continue to eject mass from the star until the final core mass is between $35M_{\odot}$ and $50M_{\odot}$, and the instability is avoided \cite{Woosley2017}.  For stars with helium-core masses in the $64M_{\odot}-133M_{\odot}$ range, the first PPI is more energetic than the star's binding energy, and the star is completely destroyed in a PISN.  Because of this, it is thought that no star can produce a BH with a mass between $46M_{\odot}$ and $133M_{\odot}$ \cite{Woosley2019}, though we note that stars formed from the mergers of other massive stars may not obey this constraint \cite[e.g.,][]{Mapelli2016}.  In our prescription for PPI/PISN, based on that developed in \cite{Belczynski2016a}, any star with a pre-collapse He-core mass between $45M_{\odot}$ and $65M_{\odot}$ is ground down to $45M_{\odot}$ by PPIs (with the BH mass reduced by a further 10\% in the conversion from baryonic to gravitational mass), while any core mass between $65M_{\odot}$ and $135M_{\odot}$ is completely destroyed in a PISN.

At the same time, \cite{Fishbach2017} showed that, because LIGO/Virgo can detect 
more massive BBHs to higher redshift, the first 6 BBH detections already 
suggested an upper bound on BH masses of $\sim 40M_{\odot}$.  They concluded that the true maximum mass of the population could be identified with less than 40 BBH detections.  We argue that the detection of a BBH with a component in the mass gap provide significant evidence of a dynamical formation history for that object.   This {has been noted previously} \cite{Fishbach2017,Gerosa2017,Rodriguez2018}, but without the cosmological model for GC formation or {varying initial spin distributions, making this work the first to produce a realistic astrophysical population of 2G BBH mergers.}  Furthermore, the detection of BBHs in the mass gap would provide information about the total contribution to the BBH merger rate from clusters, since the fraction of mass-gap BHs to the total number of mergers can be theoretically predicted.  

In Figure \ref{fig:massgap}, we show the fraction of all BBH mergers from GCs 
that are the result of multiple mergers as a function of the  initial spin of 1G 
BHs.  If it is assumed that all BHs from stars are  born with zero spin, then 
nearly $13\%$ of BBH mergers from GCs are 1G+2G mergers, while $18\%$ of all 
detected sources are.  Only $1\%$ of mergers are 2G+2G, though this contributes 
3\% of detected sources.  Of the BBH mergers with at least one 2G component, 
$9\%$ of all mergers ($7\%$ at $z<1$) have one component mass greater than 
$40.5M_{\odot}$ (representing $13\%$ of the detected population), while 6\% (3\% 
at $z<1$) have a component greater than $55M_{\odot}$ ($7\%$ of detected BBHs).  
Although the largest mass BH that can form from a single star in our simulations 
is $40.5M_{\odot}$ \cite{Rodriguez2018,Belczynski2016a}, we assume a threshold 
of $55 M_{\odot}$ as our gold-standard for identifying BHs in the mass gap.  
This is largely motivated by differences in the various population synthesis 
approaches to implementing the PPI/PISN physics \cite[e.g., 
][]{Belczynski2016a,Spera2017,2019arXiv190402821S}, though we note that the most recent supernova studies with realistic binary stellar evolution and PPI physics produce a maximum BBH component mass of $46M_{\odot}$ \cite{Woosley2019}.  

As we consider models with larger 1G BH birth spins, the fraction of 2G BBH 
mergers decrease dramatically, as the GW recoils eject significantly more of the 
1G BBH merger products from the cluster.  Increasing the birth spins of 1G BHs 
from 0.0 to 0.1 decreases the fraction of BBH mergers with a 2G component by more than a factor of two, while the fraction of detected BBH mergers with a component definitively in the mass gap ($\gtrsim 55M_{\odot}$) decreases from $7\%$ to 4\%.  If we consider a universe where $\chi_{\rm birth} = 0.5$ for all 1G BHs, less than $1\%$ of BBH mergers from globular clusters contain 2G BHs, and only $0.3\%$ of detected mergers would be definitively in the mass gap.  Of course, if the spin magnitudes of BHs were $\sim 0.5$, then the contribution to the BBH merger rate from GCs could be easily identified by spin measurements alone, \cite[e.g.][]{Rodriguez2016c,Farr2017,Farr2018}, regardless of the contribution of 2G BBH mergers.

\section{Masses and Spins}
\label{sec:massspin}

\begin{figure*}[tbh]
\centering
\includegraphics[scale=0.75, trim=0.3in 0.5in 0in 0.1in, clip=true ]{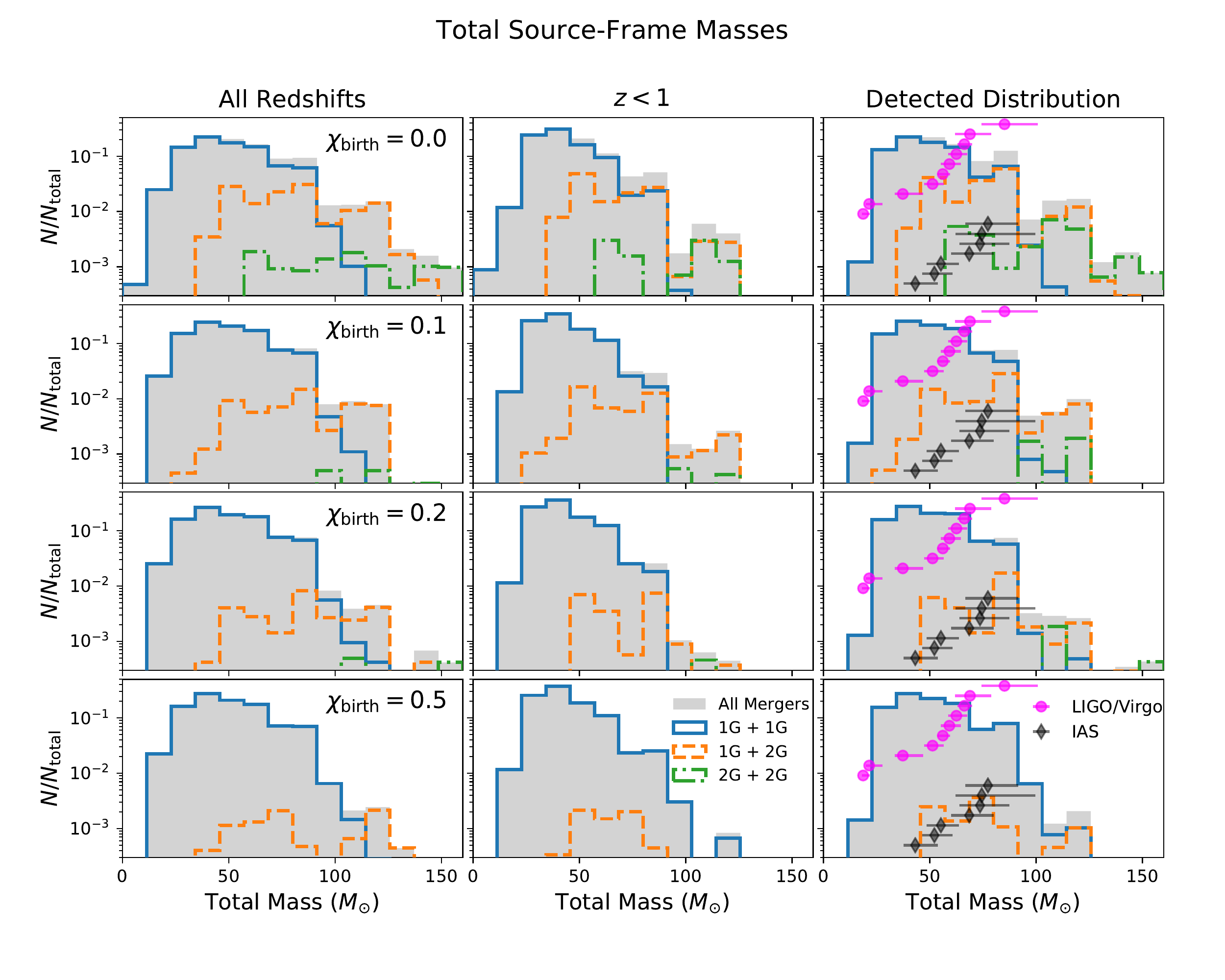}
\caption{The distribution of total BBH masses for the mergers from our four GC 
populations with different birth spins for 1G BHs.  The $\chi_{\rm birth} = 0.0$, 
0.1, 0.2, and 0.5 distributions are shown in each row from top to bottom.  The left hand column 
shows all mergers over all redshifts, while the middle column shows mergers 
occurring in the local universe ($z<1$).  The right hand column shows the 
distribution of BBHs detectable by a three-detector LIGO/Virgo operating at 
design sensitivity.   The filled grey histogram shows the distribution of all 
BBH mergers, while the solid blue, dashed orange, and dot-dashed green lines 
show the contribution from 1G+1G, 1G+2G, and 2G+2G mergers, normalized to the 
total number of BBH mergers from all generations.  In the detectable column, the 
fuchsia and turquoise ticks show the the total masses from the LIGO/Virgo and IAS catalogs from O1 and O2.}
\label{fig:masses}
\end{figure*}

We now explore the mass, mass ratio, and spin distributions of 
1G and 2G BBHs from our four cluster populations, and briefly compare them to 
the current LIGO detections.  A full statistical comparison between the 
distributions presented here and the LIGO/Virgo posterior probability 
distributions is beyond the scope of this paper, but for reference we show the 
{median and 90\% credible intervals for the masses and effective spins of the 10 LIGO/Virgo BBH 
detections.}  We also show the BBH catalog released by 
\cite{Roulet2019,Venumadhav2019,Zackay2019,Venumadhav2019a}, containing a 
reanalysis of the 10 published LIGO/Virgo events and 7 new BBHs candidates 
identified on O1 and O2 {(though we only 
show the 7 new BBH candidates that were not identified in the original 
LIGO/Virgo catalog)}.  We refer to this as 
the IAS catalog.  

\subsection{Masses}
\label{subsec:mass}

\begin{figure*}[tb]
\centering
\includegraphics[scale=0.75, trim=0.3in 0.5in 0in 0.1in, clip=true ]{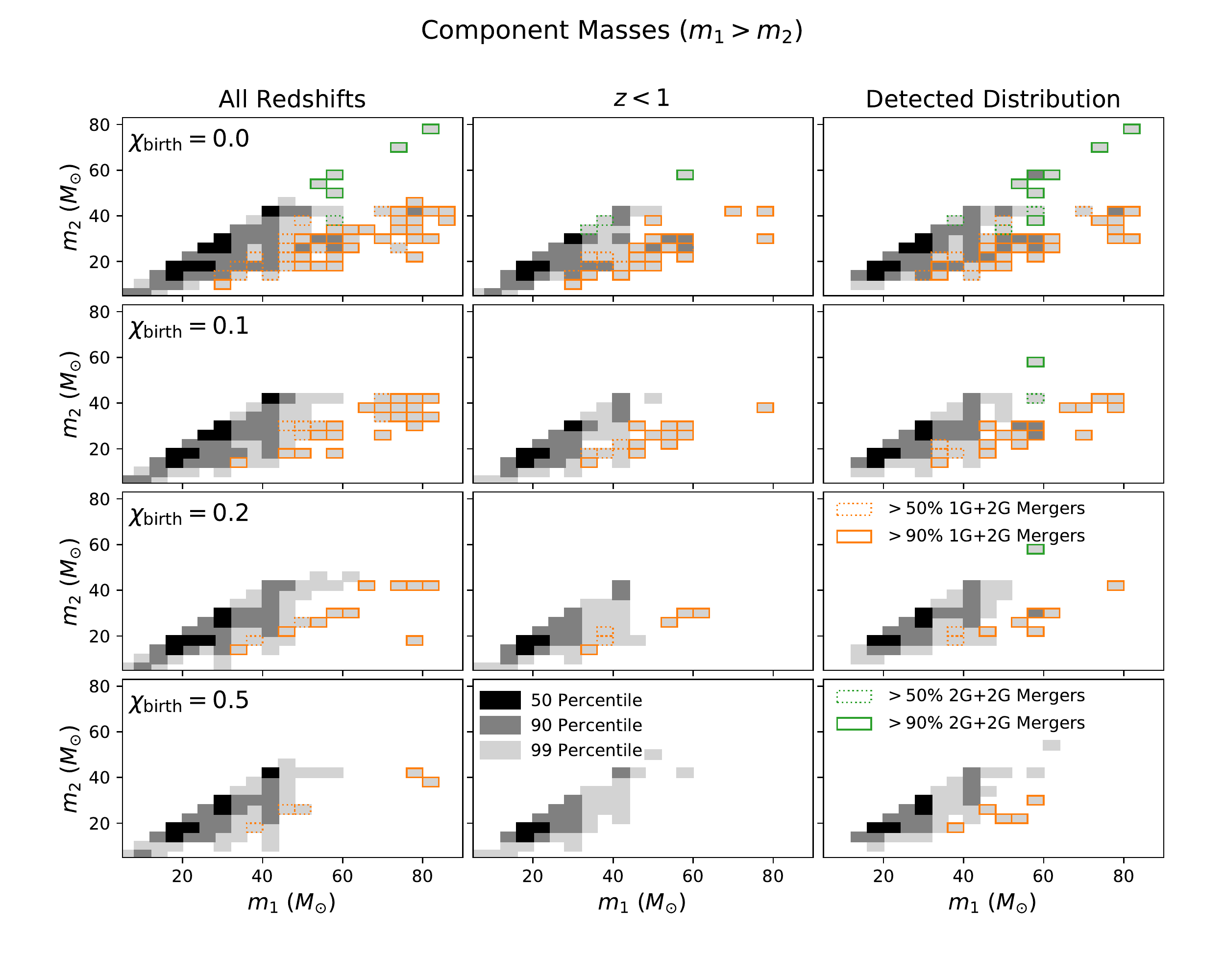}
 \caption{Similar to Figure \ref{fig:masses}, but showing the joint component 
 mass distributions from our four GC populations with different birth spins 
 divided into square bins of $4M_{\odot}$. The contours containing 50, 90, and 99 percent of all sources are shown in black, gray, and light gray, respectively.  Each bin is highlighted in dotted (solid) orange if more than 50\% (90\%) of the mergers in that bin are 1G+2G BBHs.  A similar scheme (in green) is used to indicate 2G+2G BBHs.}
 \label{fig:compmasses}
 \end{figure*}

\begin{figure*}[tb]
\centering
\includegraphics[scale=0.75, trim=0.3in 0.5in 0in 0.1in, clip=true ]{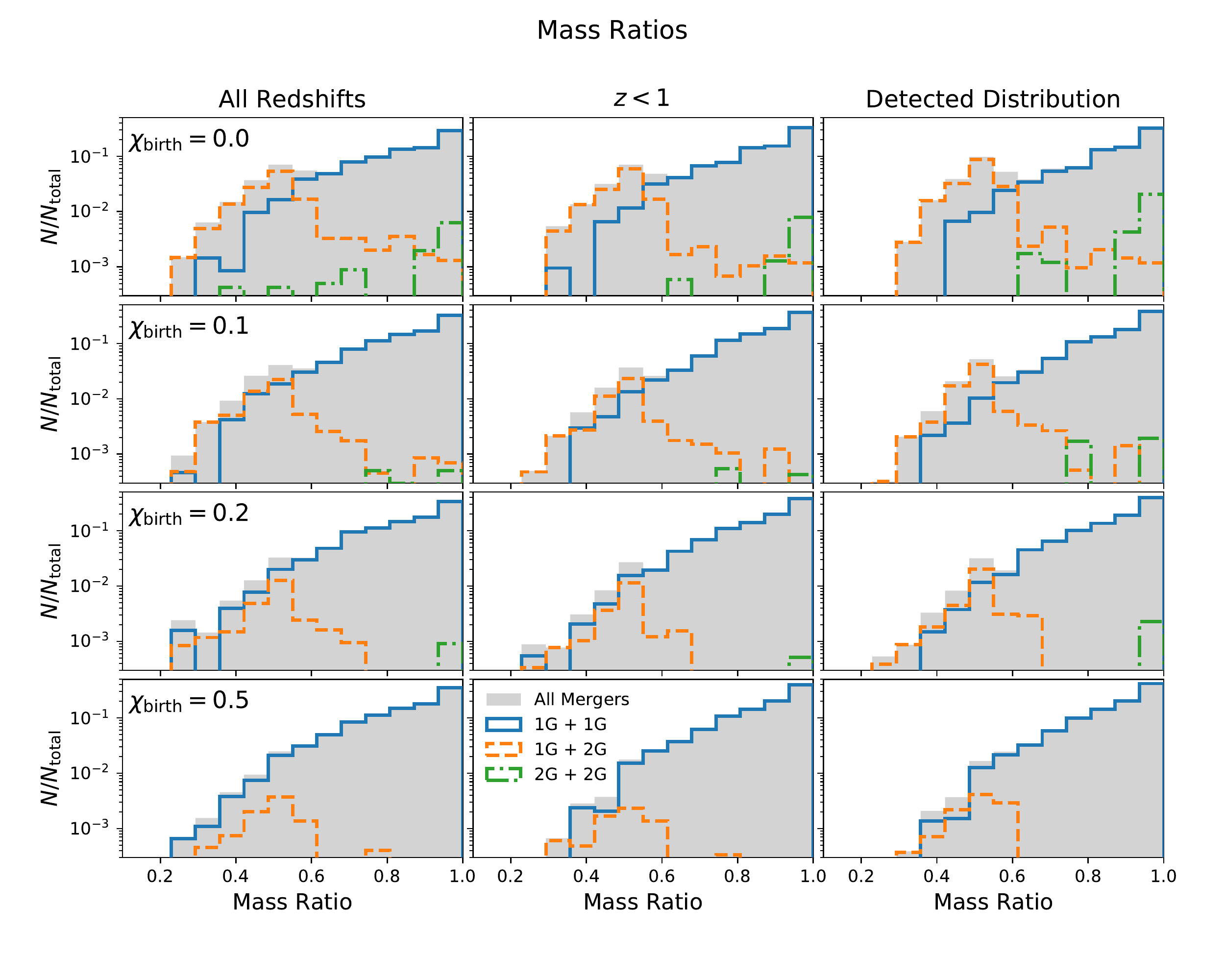}
\caption{Similar to Figure \ref{fig:masses}, but showing instead the distribution of mass ratios, $m_2 / m_1$ (where $m_2 < m_1$) for 
all BBH mergers from the four GC spin populations.  {Note that we do not show the posterior distributions for the LIGO/Virgo or IAS events, as the 90\% credible regions are very poorly constrained.}}
\label{fig:q}
\end{figure*}

In Figure \ref{fig:masses}, we show the population of total masses from each of 
our four GC populations, showing separately the distribution of BBH mergers 
across all redshifts, in the local universe ($z<1$), and the distribution 
detectable by a three-detector LIGO/Virgo network operating at design 
sensitivity.  We divide each population into 1G+1G, 1G+2G, and 2G+2G 
sub-populations, each normalized to the total BBH merger population.  As was obvious in Section \ref{sec:2gbbh} and Figure \ref{fig:massgap}, a significant population of 2G+2G mergers can only be produced from GCs in the case where the birth spins of 1G BHs are zero.  As a result, the population of BBHs with total masses $> 120M_{\odot}$ virtually disappears when $\chi_{\rm birth} > 0$.  

The population of 1G+2G BBH mergers is less dependent on the BH birth spins than 
the 2G+2G mergers, since almost any 2G BH retained in the cluster will be 
ejected as a binary.  As an example, in a cluster with $10^6$ initial particles, 
$r_v = 1$pc, and $Z=0.01Z_{\odot}$, 48 2G BHs are produced by in-cluster mergers 
of 1G+1G binaries, 31 of which are retained by the cluster when $\chi_{\rm 
birth} = 0$.  Of those 31 BHs, 16 are later ejected from the cluster as binaries 
while 11 merge again inside the cluster, but only 4 are ejected as single BHs.  
This is in stark contrast to the 1G BHs, of which nearly 76\% are ejected as 
single BHs, a fraction consistent with previous semi-analytic estimates for the 
number of single stars ejected by a single hard binary 
\cite{Goodman1984,Heggie2003}.  This difference arises because 2G BHs are 
typically the most massive BHs in the cluster at any given time, making them 
more likely to exchange into a less-massive BBH \cite{Sigurdsson1993a}.  
Additionally, such 2G BHs are less likely to be ejected from the cluster during 
an encounter with a BBH, since most 2G BHs will be similar in mass to 1G+1G BBHs.  

As the birth spin of BHs is increased, the fraction of BBHs with masses greater than $80M_{\odot}$ also decreases significantly in Figure \ref{fig:masses}.  Somewhat surprisingly, there exists a small population of 1G BBHs with total masses $\sim 120M_{\odot}$, even when $\chi_{\rm birth} = 0.5$.  These handful of objects, while rare, are produced very early in the cluster lifetime, either by stable mass transfer onto a $40.5M_{\odot}$ BH, or by stellar mergers which produce massive stars with atypically large hydrogen envelopes and small helium cores \cite[][]{Mapelli2016,DiCarlo2019}.  These objects are largely the result of the stellar merger handling in BSE and our adopted PPI/PISN prescriptions, and it is not obvious whether such objects could exist in nature.  

At first glance, it appears that the lack of GW sources with total masses 
$\gtrsim 100M_{\odot}$ would suggest against a universe where $\chi_{\rm birth} 
= 0$ for all 1G BHs.  However, we stress that, with 10 BBH detections from 
LIGO/Virgo (and 17 candidates from the IAS), the lack of such super-heavy BBHs 
is still consistent with the statistics quoted here, since only 9\% of detected 
BBHs have total source-frame masses greater than the most massive BBH identified 
to date \cite[GW170729, at $85M_{\odot}$, 
][]{TheLIGOScientificCollaboration2018a}, and only $4\%$ of detected BBHs have 
total masses greater than $100M_{\odot}$, even when $\chi_{\rm birth} = 0.0$.  
These fractions decrease to 2\% above $85M_{\odot}$ and 0.5\% above 
$100M_\odot{}$, respectively, when $\chi_{\rm birth} = 0.5$.  However, as the size of the LIGO/Virgo BBH catalog continues to grow, it will become easier to either identify (or rule out the existence of) such massive BBHs in the universe.

In Figure \ref{fig:compmasses}, we break down the events presented in Figure 
\ref{fig:masses} into their individual components.  Instead of showing full 2D 
histograms, for simplicity we only show where 50\%, 90\%, and 99\% of all 
sources lie in the $m_1$-$m_2$ plane.  We also show which bins are dominated by 
1G+2G and 2G+2G mergers.  There it becomes obvious that any significant number 
of detections, certainly within the 90$^{\rm th}$ percentile, cannot be produced 
with component masses above $40M_{\odot}$ when the birth spins of BHs are large.  We also note that GW170729 can be easily formed in the $\chi_{\rm birth}=0.0$ models, where it lies in the region and mass bin that is dominated (more than 50\% of mergers) by 2G BHs, although many of the BBHs in that region are also composed of 1G BHs.  This is consistent with statistical studies of GW170729 \cite{Kimball2019,Chatziioannou2019}, suggesting that while the event is consistent with a 2G BBH merger, there is insufficient evidence to say definitively.  

Finally, we note that the two lowest mass BBHs in Figures \ref{fig:masses} and 
\ref{fig:compmasses} also appear to be just outside the 99 percentile regions 
for the masses from GCs.  This is because we have restricted ourselves to 
classical GCs with large virial radii and low metallicities.  However, by 
considering systems with $Z\sim Z_{\odot}$, such as super-star clusters or open 
clusters \cite[e.g.,][]{Ziosi2014},  dynamics can easily produce such low-mass 
BBHs \cite{Chatterjee2017a}.  Furthermore, the current collection of GC models 
do not form low-mass BBHs because any 5-10$M_{\odot}$ BHs that remain in the 
cluster are not participating in dynamical encounters at the present day.   
BBH-forming encounters are dominated by the most massive remaining BHs in the 
clusters \cite[][]{Morscher2015}, which with our assumed initial conditions are typically in the 10-15$M_{\odot}$ range.  However, if GCs are born with significantly smaller initial virial radii ($\sim 0.5$pc), then all BHs, including the low-mass BHs, will be processed into binaries and ejected from the cluster by the present day.  Such ultra-compact clusters are not considered here, but are necessary to explain the observation of core-collapsed GCs in the Milky Way and other galaxies \cite{Kremer2018,Kremer2019}.  

\subsection{Mass Ratios}

The dynamical formation of BBHs in GCs typically involves the most massive BHs 
available in the cluster at any given time \cite[e.g,][]{Morscher2015}.  Even if 
a BBH were to form with a significantly low mass ratio, repeated binary-single 
and binary-binary encounters would preferentially exchange BHs into the binary in 
favor of creating a nearly equal-mass system \cite{Sigurdsson1993a}.  The BBHs that 
merge in the cluster therefore tend to have nearly equal mass components drawn from the 
most massive BHs in the cluster.  When a BBH merges, its 2G merger product --- if it is retained in the cluster --- is 
then nearly twice the mass of the most massive 1G BHs.  And because GCs 
typically only harden $\sim 1$ BBH at any given 
time, that 2G BH is most likely to rapidly merge again before the cluster can 
form another 2G BH.  We note that this is not true in 
NSCs, where the larger escape speeds make it possible to retain 
mergers of even 2G BBHs, potentially building several successive generations of 
BBH mergers \cite[][]{Antonini2016, 2019arXiv190605295G}.

In Figure \ref{fig:q}, we show the mass ratio distributions for our four 
$\chi_{\rm birth}$ populations of GCs.  As expected, the distribution of 1G+1G 
BBHs piles up strongly at a mass ratio of 1, as seen in previous dynamical 
studies \cite{Rodriguez2016b}.  However, the 1G+2G BBHs peak at a much lower 
mass ratio of $\sim 0.5$, because the 2G BH in these binaries is typically twice 
the mass of the most massive 1G BHs in the cluster.  The detected distribution 
of BBHs shows a significant secondary peak in the mass ratio distribution at 
$q\sim 0.5$, driven by the more massive 1G+2G and their correspondingly larger 
detection weights.  Finally, the handful of 2G+2G BBHs typically have mass 
ratios closer to unity.  This is consistent with the trend towards equal mass 
binaries: if a cluster manages to retain two 2G BHs at once, one 2G BH will 
likely eject any 1G BH bound to its fellow 2G BH, in favor of creating a 
near-equal mass 2G+2G system.


\subsection{Effective Spins}
\label{sec:spins}


\begin{figure*}[tb]
\centering
\includegraphics[scale=0.75, trim=0.3in 0.5in 0in 0.1in, clip=true ]{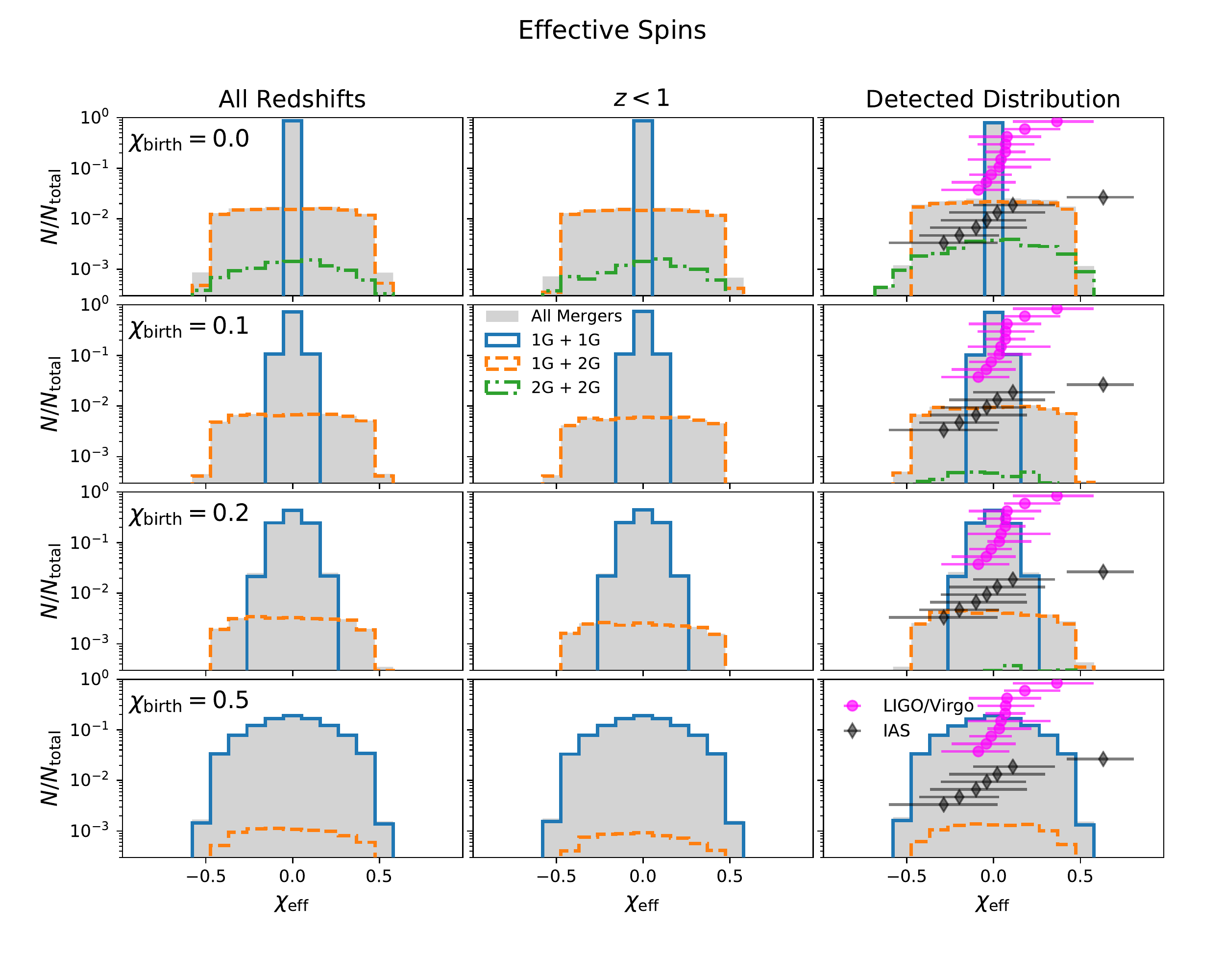}
\caption{Similar to Figure \ref{fig:masses}, but showing instead the distribution of effective spins for 
all BBH mergers from the four GC spin populations.  Note that we have reweighted the posterior distributions from the IAS catalog to use a prior probability distribution with uniform spin magnitudes and isotropic spin
orientations.}
\label{fig:chieff}
\end{figure*}

Even for non-spinning BHs, a BBH merger produces a remnant with $\chi \sim 0.7$ \cite[e.g.,][]{Berti2007,Tichy2008,Lousto2010}.  Because of this, we expect BBH mergers with 2G components 
to have some spin regardless of the $\chi_{\rm birth}$ of 1G BHs.  Unfortunately, what GW experiments measure best is not the individual spins of the components, but the effective spin of the BBH, $\chi_{\rm eff}$, given by the mass-weighted projection of the spins 
onto the orbital angular momentum of the binary:

\begin{equation}
	\chi_{\rm eff} = \left[\frac{m_1 \vec{\chi_1} + m_2 \vec{\chi_2}}{m_1+m_2} 
	\right] \cdot \hat{L}~.
	\label{eqn:chieff}
\end{equation}

\noindent For dynamically-assembled BBHs, the angle between the spin and 
orbital angular momenta is expected to be isotropically distributed 
\cite{Rodriguez2016c}, suggesting that the distribution of $\chi_{\rm eff}$ 
should be symmetric and centered on zero, with a tail determined by the spins
of the components and the mass ratio distribution of the binaries.


In Figure \ref{fig:chieff}, {we show the $\chi_{\rm eff}$ distributions for our four $\chi_{\rm birth}$ populations}.   If we assume that 
LIGO/Virgo can {confidentially exclude non-zero} spins for BBHs with $|\chi_{\rm eff}| \gtrsim 0.2$ 
\cite[e.g.,][]{Vitale2017,TheLIGOScientificCollaboration2018a}, then the worst-case scenario for detecting the spin of BBHs from 
dense star clusters is
the case where $\chi_{\rm birth} = 0.2$.  There, only 1\% (2\%) of the 
actual (observed) distribution of BBHs will merge with $|\chi_{\rm eff}| > 0.2$.  
If 1G BHs are born with no spin, the production of 2G BBH mergers through 
repeated mergers can produce a population with significant spin, with 8\% 
(11\%) of the actual (observed) population of BBHs having $|\chi_{\rm eff}| > 0.2$.  
On the other hand, if the spins of 1G BHs are 0.5, then 37\% of all
BBH mergers (actual and observed) will have $|\chi_{\rm eff}| > 0.2$.    
Because we have assumed that all 1G BHs are born with the 
same spins regardless of their masses, there is no difference between the 
observed and actual distributions of $\chi_{\rm eff}$ for 1G BBH mergers.  
However, because BBHs with 2G components are characteristically more massive, 
they are detectable in a larger volume of space.

{As with Figure \ref{fig:masses}, we also show the median and 90\% regions of the 1-dimensional marginalized posteriors for $\chi_{\rm eff}$ in Figure \ref{fig:chieff}.  However, the posterior distributions provided in \cite{Venumadhav2019a} and \cite{TheLIGOScientificCollaboration2018} were computed using different prior probabilities for the spin distributions.  The LIGO/Virgo parameter estimation prior assumes a uniform distribution of component spin magnitudes with the spin orientations isotropically distributed on the sphere.  The IAS parameter estimation, on the other hand, uses a uniform prior on $\chi_{\rm eff}$.  This flat prior distribution is not peaked at $\chi_{\rm eff} =0 $ (unlike the LIGO/Virgo prior) and is partially responsible for the large spins --- e.g.~the median $\chi_{\rm eff}$ of 0.81 and $-0.7$ for GW151216 and GW170403 respectively --- reported by the IAS analysis.  To present our results self-consistently, we reweight the posterior samples from the IAS using the same prior probability distribution employed by LIGO/Virgo analysis.}

{Unlike the previous plots, the distribution of measured $\chi_{\rm eff}$ shown 
in Figure \ref{fig:chieff} is significantly different between the LIGO/Virgo and 
IAS catalogs.  As stated above, the IAS catalog contains a candidate BBH merger 
--- GW151216, with median posterior probability of $\chi_{\rm eff} = 0.81$ (under the flat $\chi_{\rm eff}$ prior) or $\chi_{\rm eff} = 0.63$ (under the uniform in spin magnitude/isotropic in spin direction prior)  --- that cannot be easily produced by any of the models presented here \cite{Zackay2019,Venumadhav2019}.  We find no BBH models with $\chi_{\rm eff} = 0.81$, and even if all 1G BHs were born with maximal spins \cite[an assumption already disfavored by model selection of 
the 10 LIGO/Virgo BBHs,][]{TheLIGOScientificCollaboration2018}, less than 2\% of 
BBH mergers would have $\chi_{\rm eff} > 0.81$. If we instead assume a prior with isotropic spin directions and uniform spin magnitudes, we can produce 2G+2G mergers with $\chi_{\rm eff} = 0.63$, but even then, only 0.2\% of detected 2G+2G mergers have sufficiently aligned spins in the $\chi_{\rm birth} = 0.0$ model.  We note that GW151216 
is only given a 71\% of being of astrophysical origin \cite{Venumadhav2019a}; however, if 
further detections reveal these events to be part of a population of highly 
spinning and aligned BBHs, it would suggest another formation mechanism for BBHs 
may also operate in the universe, such as the chemically-homogeneous binary evolution channel \cite{DeMink2016,Mandel2016a,Marchant2016}.}

{While cluster dynamics cannot easily produce the most highly spin-aligned BBH candidate, it is a natural explanation for the most spin-anti-aligned event, GW170121.  This event from the IAS catalog, which (unlike GW151216) has a $>$99\% chance of being astrophysical, is also the first BBH event with significant spin anti-alignment.  Under both priors, 90\% of the posterior support for $\chi_{\rm eff}$ is less than zero.  Furthermore, the median value of the $\chi_{\rm eff}$ posterior ($-0.3$ when a flat prior is employed, and $-0.2$ when using an isotropic/uniform spin prior).  This configuration is unlikely to arise from isolated binary evolution without significant BH natal kicks \cite{Rodriguez2016c,2018PhRvD..98h4036G}, but can easily be explained by dynamical processing of BBHs.  6\% of all detected BBHs in our $\chi_{\rm birth} = 0$ models have $\chi_{\rm eff} \leq -0.2$.  As the birth spins are increased, this number decreases to 1\% of all mergers when $\chi_{\rm birth} = 0.2$ (due to the depletion of 2G BHs) before increasing to more than 18\% of detected mergers when $\chi_{\rm birth} = 0.5$.  If we assume that GW170721 originated in a dense stellar environment, then its large, negative effective spin would require either significant birth spins for 1G BHs or significant numbers of 2G BHs (and correspondingly low birth spins).  However, models with $\chi_{\rm birth} = 0.1$ or 0.2 would be less likely to produce such systems, as only 2\% and 1\% of detected BBHs from those models have $\chi_{\rm eff} \leq -0.2$.}


\subsection{Spin Magnitudes}
\label{sec:spinmag}

\begin{figure}[tbh]
\includegraphics[scale=0.72, trim=0.1in 0.0in 0in 0.0in, clip=true]{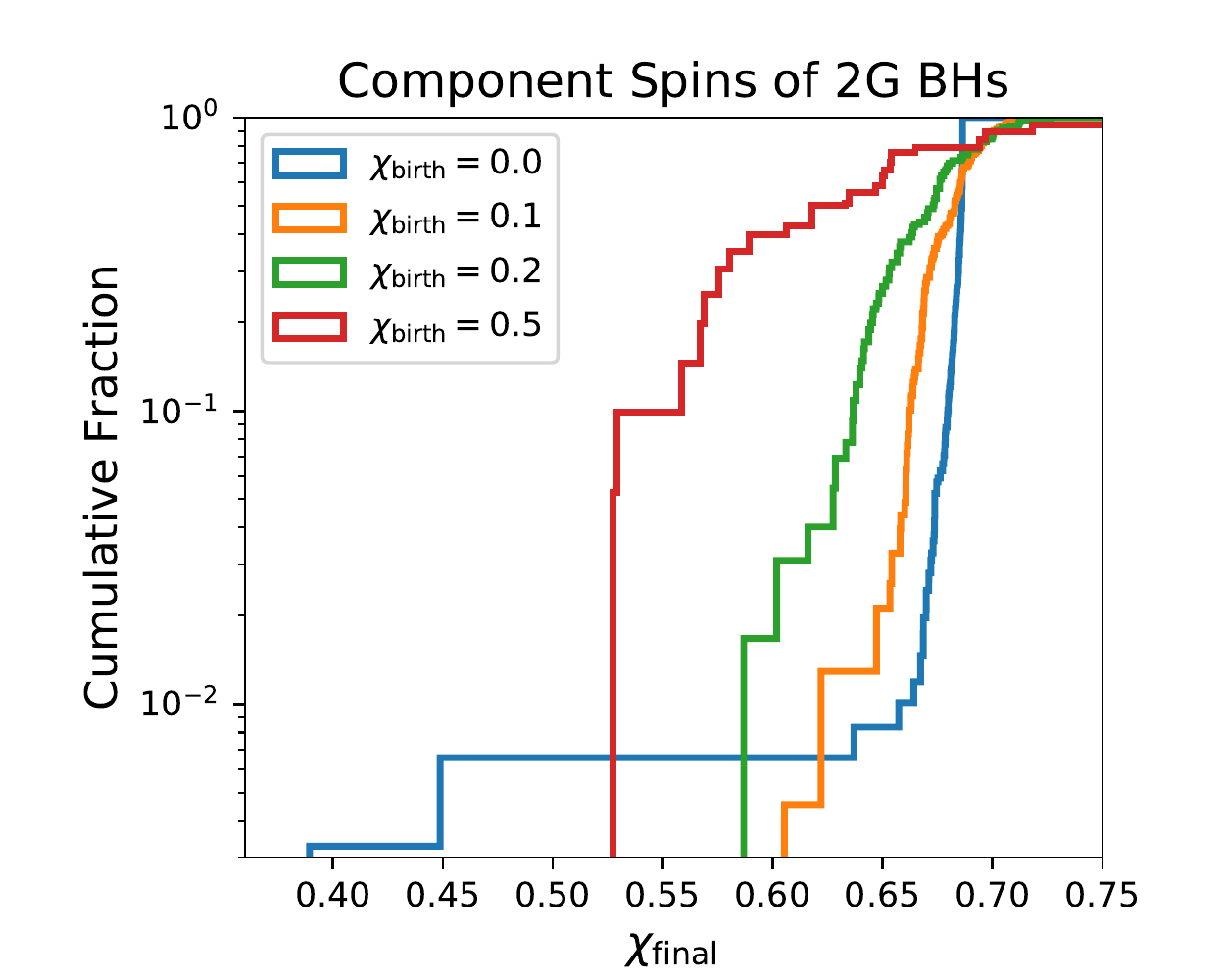}
\caption{The cumulative distribution of the component spins for 2G BHs which merge across all redshifts.  We show the distributions for our $\chi_{\rm birth} = 0$, 0.1, 0.2, and 0.5 populations in blue, orange, green, and red, respectively.  As the birth spins of BHs are increased, the distribution broadens from being very strongly peaked at $\chi_{\rm final}\approx 0.69$ when $\chi_{\rm birth}= 0.0$ to having a median of 0.62 when $\chi_{\rm birth}= 0.5$.  For completeness, we also include the two 3G BHs that form and merge (with lower $\chi_{\rm final}$) in the $\chi_{\rm birth}= 0.0$ population; see section \ref{subsec:model}.}
\label{fig:compspin}
\end{figure}

The merger of two non-spinning, equal-mass BHs will produce a final BH with a spin of $\chi_{\rm final} \sim 0.69$.  However the final spin of the newly-formed BHs depends strongly on both the mass ratio, with smaller mass ratios preferring lower spins \cite[e.g.,][]{Berti2007}, and the spin magnitudes and orientations at the point of merger \cite[][]{Rezzolla2008,Buonanno2008,Tichy2008}.  Of course, while this 7-dimensional parameter space (the two BH spin vectors and the mass ratio) determine the final spin of the BH, they are also the parameters responsible for the magnitude of the GW recoil kicks imparted to BBH merger remnants.  Because of this, certain regions of final mass and spin parameter space for 2G BBHs are inaccessible in realistic clusters, because the BBH configuration required to produce a given remnant would also result in a GW kick greater than the escape speed of the cluster \cite[e.g.,][]{Merritt2004}.  

As an example, in our $\chi_{\rm birth} = 0.0$ models, the entirety of the GW 
recoil kick for 1G+1G BBH mergers is driven by the mass ratio of the system: 
when the components are of equal mass, no kick is given, and the spin of the 
remnant is $\chi_{\rm final}\approx 0.69$.  As the mass ratio is decreased, the 
kick increases, to a maximum of $\sim 175\pm11 \rm{km}/\rm{s}$ when the mass ratio is 
$q\sim 1/3$ \cite{Gonzalez2007}, while the remnant spin decreases to $\chi_{\rm final}\approx 0.55$, 
\cite[c.f.~Table 1 of][]{Berti2007}. In Figure \ref{fig:compspin}, we show the 
spin magnitudes of all the 2G BBHs that merge in our four GC universes, and can 
clearly see that the distribution of 2G spin magnitudes from $\chi_{\rm 
birth}=0$ universe are strongly concentrated at $\chi_{\rm final}\approx 0.69$, 
with 90\% of sources lying between 0.67 and 0.69.  As the birth spin of BHs is 
increased, the component spins of the 2G BHs become less concentrated: the 
distribution of final spins for the $\chi_{\rm birth}=0.5$ universe has a median 
of 0.62, with 90\% of sources having final spins between 0.53 and 0.72.  This 
broadening is largely a result of the increasing parameter space of BBH mergers 
that produce low-kick BBH mergers as we consider systems where the spin vectors 
of the two BHs are important. We leave a full mapping of the final BH spin and BH retention fraction to future work.

\section{Conclusions}

In this paper, we explored the production, merger properties, and detectable 
populations of 2G BHs: BHs which were forged by previous BBH mergers in the 
cores of dense star clusters.  Using self-consistent dynamical models of GCs 
with different birth spins for BHs, we showed that if all BHs created from 
stellar collapse are born with no spin, then more than 10\% of all BBH mergers from clusters (and nearly 20\% of the detections) should have at least one component created during a previous merger.  Of those, $\sim 7\%$ would have at least one component above $55M_{\odot}$, placing it clearly in the upper-mass gap where the formation of BHs from single or binary stars is inhibited by the PPI/PISN mechanism.  If the birth spins of 1G BHs is higher, the retention of 2G BHs by the cluster decreases, and the number of 2G BBH mergers drops precipitously.  In the largest spin model we consider, where $\chi_{\rm birth} = 0.5$, less than 3\% of 2G BHs are retained by the cluster, and less than 1\% of detectable BBH mergers contain 2G BHs.     

As previously stated, if all 1G BHs were born with spins of $\chi_{\rm 
birth}=0.5$, then measurements of the spins would themselves be an 
effective tool to distinguish BBH formation scenarios.  As GW parameter estimation can 
reliably measure $\chi_{\rm eff}$ to within a 90\% uncertainty of $\pm 0.2$, we would 
expect that LIGO/Virgo would already have evidence for or against the 
dynamical formation scenarios (as $\sim 20\%$ of all BBHs from GCs would have 
$\chi_{\rm eff} < -0.2$).  In many ways, the worst case scenario for identifying BBHs from the dynamical formation channel would be the case where $\chi_{\rm birth} \sim 0.2$.  In that regime, the birth spins of 1G BHs are too 
low to be reliably measured (less than $2\%$ of detected BBH mergers would have 
$\chi_{\rm eff} < -0.2$), while $\lesssim 5\%$ of detected mergers would have 
components in the mass gap.  {However, the recently identified BBH candidate GW170121 has $\chi_{\rm eff} < 0$ at 90\% confidence, regardless of the assumed prior distribution.  If it is assumed that this BBH was created in a dense stellar environment, then it suggests either that all 1G BHs are born with low spins, creating a population of highly spinning, retained 2G BBH mergers, or that the birth spins of 1G BHs are significantly larger than $\chi\sim 0.2$. }

Throughout this paper, we have limited ourselves to 2G BHs that were created 
from the mergers of previous BHs.  However, it has been proposed for many years 
that massive BHs, and even the progenitors of intermediate-mass BHs (IMBHs), 
could be forged by the repeated mergers of massive stars during the early stages 
of cluster evolution 
\cite{Zwart2004a,Zwart2002,2004ApJ...604..632G,2006MNRAS.368..141F,2006ApJ...640L..39G,Mapelli2016,DiCarlo2019}. We 
identified a handful of these objects in Section \ref{subsec:mass}, and noted 
that some merged with components $\gtrsim 55M_{\odot}$, even though they were 
considered 1G BHs in our simulations.  These objects could have important 
implications for the formation of BHs in the mass gap and for the creation of 
both IMBHs and the seeds of super-massive BHs.  However, significant work 
remains to be done to better understand the evolution of massive stars that are 
created from the mergers of other massive stars.  

We have also only considered GCs that are born with initial binary fractions of 
10\%.  This is a standard choice in stellar dynamics, as it has been shown to 
reproduce the binary fraction of present-day GCs 
\cite{Hurley2007,Chatterjee2010}.  However, given the short lifetimes of massive 
stars, the initial binary fraction of massive stars in young GCs is essentially 
unconstrained \cite[though observations in the local universe suggest fractions 
as high as 70\%; see][]{2012MNRAS.424.1925C}.  This has been shown to effect the merger rate of BBHs from GCs \cite{Chatterjee2017a}, and would have a significant impact on the production of 2G BBHs in GCs, particularly if the majority of massive stellar binaries produced BBHs that merged early in the cluster lifetime  \cite[e.g.,][]{Morawski2018}.  These 2G BHs might have distinct properties from those studied here, most of which were created through dynamical encounters.  We leave a proper study of the effects of massive binary stars and their resultant BBH mergers in clusters to future work.

Finally, we note that showing the distributions of the median values from both 
the LIGO/Virgo and IAS catalogs is a crude way to compare the results from O1 
and O2 to the distributions presented here.  A more appropriate comparison 
between different formation channels and the LIGO/Virgo results, such as those 
presented in 
\cite{Zevin2017,2017PhRvD..96b3012T,2018ApJ...856..173T,2019MNRAS.484.4008G,2019arXiv190511054B,2019MNRAS.tmp.1874P,2018PhRvD..98h3017T}, is beyond the scope of this paper.  However the true scientific potential of GW astronomy will depend on doing a proper comparison between the full 15-dimensional posterior distributions for GW events and multiple theoretical distributions, such as the ones presented here.  A study comparing the observational and theoretical distributions using Bayesian model selection techniques is currently underway \cite{Zevin2019}.

CR is supported by a Pappalardo Postdoctoral Fellowship at MIT.  This work was
supported by NASA Grant NNX14AP92G and NSF Grant AST-1716762 at Northwestern
University.  PAS acknowledges support from the Ram{\'o}n y Cajal Programme of
the Ministry of Economy, Industry and Competitiveness of Spain and the
COST Action GWverse CA16104.  CR and MZ thank the Niels Bohr
Institute for its hospitality while part of this work was completed, and
the Kavli Foundation and the DNRF for supporting the 2017 Kavli
Summer Program.  CR and FR also acknowledge support from NSF Grant PHY-1607611 to the Aspen
Center for Physics, where this work was started.


\end{document}